# When Is Molecular-Dynamics-Predicted Ionic Conductivity Reliable in Solid Electrolytes?

*Yiwei You*[1,2], *Shaofei Chen*[1], *Zhifeng Wu*[1], *Pushun Lu*[1], *Eric Jianfeng Cheng*[3], *Songyan Chen*[1], *Shunqing Wu*[1,*]

[1]Department of Physics, Key Laboratory of Low Dimensional Condensed Matter Physics (Department of Education of Fujian Province), Xiamen University, Xiamen 361005, China

[2]Institute of Technology for Carbon Neutrality, Shenzhen Institutes of Advanced Technology, Chinese Academy of Sciences, Shenzhen, 518055, China

[3]Advanced Institute for Materials Research (AIMR), Tohoku University, Sendai, 980-8577, Japan

Corresponding E-mail: wsq@xmu.edu.cn

**Abstract**

Molecular dynamics is widely used to predict ionic conductivity in solid electrolytes, but the reliability of these predictions is often difficult to assess. Our analysis identifies finite trajectory length, limited cell size, and insufficient sampling as intrinsic limitations of finite atomistic ion-transport calculations. Cubic $Li_7La_3Zr_2O_{12}$ is used as a representative solid electrolyte to quantify their consequences. These limitations can remain hidden behind apparently linear mean-squared displacements and well-behaved Arrhenius relations, leading to inaccurate diffusivities, activation energies, and extrapolated ionic conductivities. Such inaccuracies can misrank candidate solid electrolytes and consequently misdirect computational screening and experimental validation. Here, we establish the local mean-squared-displacement exponent, $\alpha(t)$, as a quantitative reliability criterion that links dynamical convergence to errors in diffusivity and Nernst-Einstein ionic conductivity. The criterion further determines the minimum trajectory length required to achieve a prescribed accuracy as a function of temperature. Independent replicas reduce statistical uncertainty, while selective single-axis expansion mitigates finite-size effects. By establishing when simulated ionic conductivity is quantitatively trustworthy, this approach enables more reliable materials ranking and more efficient use of computational and experimental resources, thereby accelerating the development of high-performance solid electrolytes for solid-state batteries.

## 1. INTRODUCTION

High ionic conductivity is a primary criterion for evaluating solid electrolytes for solid-state batteries, and molecular dynamics is increasingly used to predict this property before extensive experimental optimization[1-3]. The calculated conductivity is often used not only to characterize ion transport but also to compare different chemistries, rank candidate materials, and identify compositions worthy of further synthesis and optimization[1, 4]. An inaccurate prediction can therefore do more than introduce numerical uncertainty: it may misclassify a promising electrolyte as a poor conductor, overestimate the potential of an inferior candidate, or distort composition-transport relationships used for materials design. Such errors can ultimately misdirect computational screening and experimental effort, reducing the efficiency of solid-electrolyte discovery. Establishing when a molecular-dynamics-predicted ionic conductivity is quantitatively reliable is therefore essential for using atomistic simulations as a meaningful tool for materials evaluation and development.

In atomistic simulations, ionic conductivity is commonly estimated from the long-time tracer diffusivity obtained from the mean-squared displacement (MSD), often through the Nernst-Einstein relation when ion-ion cross-correlations are neglected or treated separately[5-8]. Because long-range hopping becomes increasingly difficult to sample near room temperature, elevated-temperature simulations followed by Arrhenius extrapolation are widely used for solid electrolytes[5, 6]. This procedure places stringent demands on temporal and spatial sampling. Finite trajectories may contain too few statistically independent hopping events and can retain contributions from pre-diffusive motion, making the extracted diffusivity sensitive to trajectory length and fitting protocol[9-12]. Finite periodic cells introduce an additional source of uncertainty because the calculated transport properties can depend on system size and the restricted dynamical modes accessible under periodic boundary conditions[6, 13, 14]. Importantly, these limitations may remain hidden behind an apparently linear MSD or a well-behaved Arrhenius relation[5, 11, 12]. Such conventional indicators are therefore insufficient, by themselves, to establish whether a predicted ionic conductivity has reached the accuracy required for quantitative materials comparison.

Previous studies have examined statistical uncertainty and finite-time estimation of diffusion coefficients[15, 16], finite-size effects under periodic boundary conditions[6, 13, 14, 17], and local MSD-based descriptions of diffusive and non-diffusive dynamics[11, 18-20]. These advances address important individual aspects of diffusion analysis, but they do not directly answer a practical question central to solid-electrolyte calculations: how accurately does a finite trajectory represent the ionic conductivity, and how much additional simulation is required to reach a desired accuracy? In particular, a quantitative connection between the local dynamical state, the resulting error in ionic transport, and the required simulation time would turn dynamical convergence from a qualitative diagnostic into an actionable criterion for materials evaluation.

In this paper, cubic $Li_7La_3Zr_2O_{12}$ (LLZO) is used as a representative solid electrolyte to quantify these general limitations of finite atomistic ion-transport simulations. We first resolve the distinct contributions of finite trajectory length, limited cell size, and statistical sampling to the calculated diffusivity. We then establish the local MSD exponent as a quantitative reliability criterion that connects dynamical convergence to errors in diffusivity and Nernst-Einstein ionic conductivity, and use this relation to determine the minimum trajectory length required for a prescribed accuracy as a function of temperature. Finally, independent replicas and selective single-axis expansion are evaluated as computationally economical strategies for reducing statistical uncertainty and finite-size effects, respectively. By making the accuracy of simulated ionic conductivity explicitly assessable, this approach helps ensure that computational screening reflects intrinsic ion-transport performance rather than numerical limitations, thereby improving the efficiency and reliability of solid-electrolyte discovery and development for solid-state batteries.

## 2. RESULTS

### *2.1 Finite Simulation Conditions Can Mask Errors in Predicted Ionic Conductivity*

In molecular dynamics simulations, the tracer diffusion coefficient is obtained from the long-time slope of the mean-squared displacement (MSD),

$$D = \frac{1}{2d} \lim_{t\to\infty} \frac{d\,\mathrm{MSD}(t)}{dt}, \qquad (1)$$

where $d$ is the dimensionality. When ion–ion cross-correlations are neglected, $D$ is related to the Nernst-Einstein ionic conductivity by

$$\sigma_{\mathrm{NE}} = \frac{nq^2D}{k_{\mathrm{B}}T}, \quad \frac{\Delta\sigma_{\mathrm{NE}}}{\sigma_{\mathrm{NE}}} = \frac{\Delta D}{D}, \qquad (2)$$

where $n$ and $q$ are the number density and charge of the mobile ions, respectively[7, 8]. Thus, an error in the calculated diffusivity is transferred directly to the predicted conductivity under otherwise identical conditions. The key requirement is therefore that the finite trajectory actually reaches the long-time transport regime represented by **Eq. 1**.

**Fig. 1** shows that this requirement is not automatically satisfied. At 300 K, the calculated $\ln D$ decreases systematically as the LLZO cell is enlarged from approximately 13 to 52 Å, whereas further expansion produces only minor changes (**Fig. 1a**). The spread among independent estimates also narrows with increasing cell size. A similarly pronounced dependence appears with simulation time. Diffusivities obtained using fitting windows of 5-300 ps remain relatively close at high temperature but increasingly diverge on cooling, while the extended 300 K trajectory approaches a different long-time value (**Fig. 1b**). At 300 K, several apparently reasonable fitting windows likewise yield substantially different diffusivities before convergence is reached (**Fig. 1c**). Consequently, a smooth MSD or an apparently linear Arrhenius relation cannot, by itself, establish the reliability of the resulting ionic conductivity.

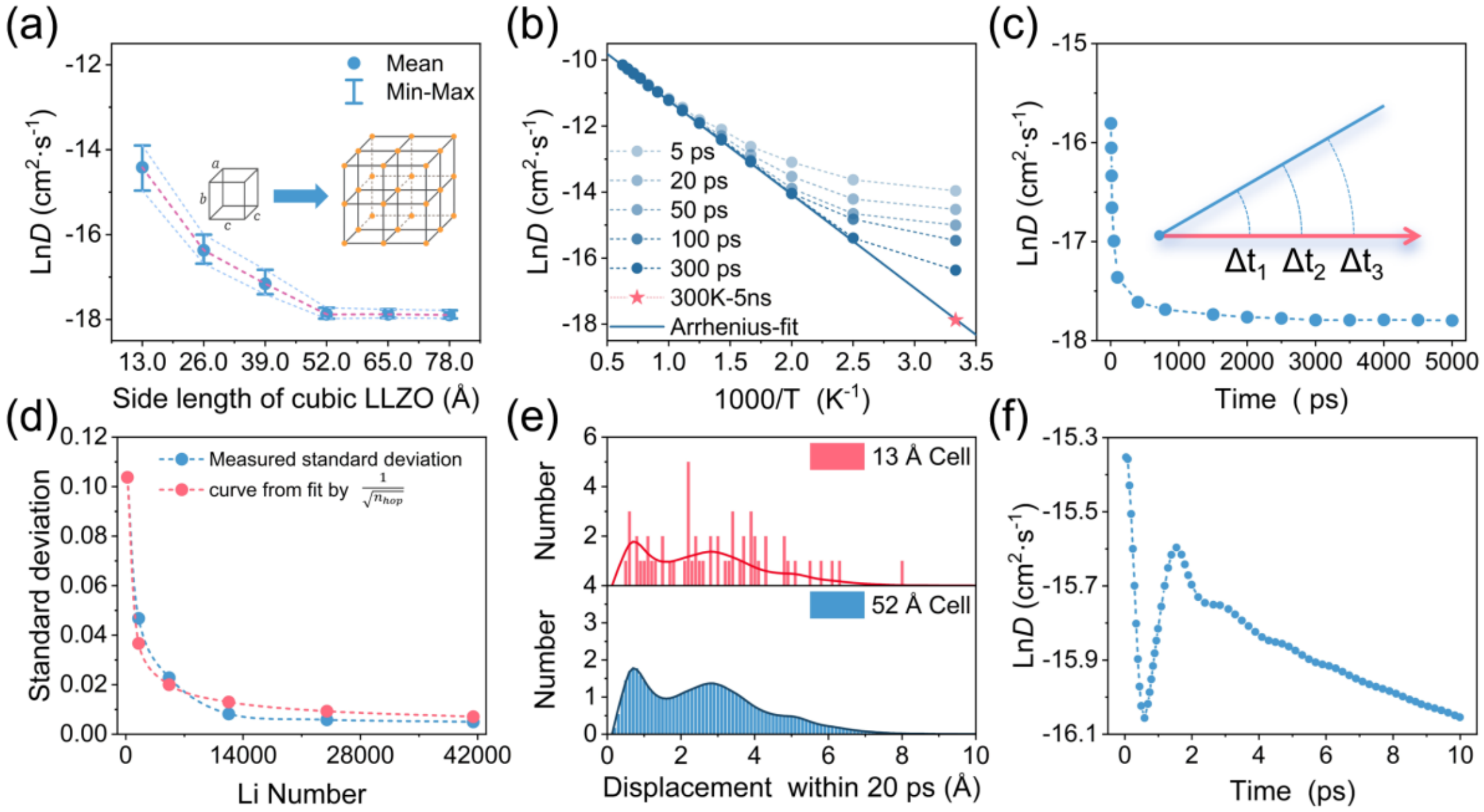

**Fig 1. Finite-simulation effects on Li-ion diffusivity underlying ionic-conductivity prediction.** (a) Mean and minimum-to-maximum range of *ln D* as a function of cubic LLZO cell size at 300 K. (b) Arrhenius relations obtained using fitting windows of 5-300 ps, together with the extended 300 K trajectory. (c) Dependence of *ln D* on fitting-window length at 300 K. (d) Standard deviation of *ln D* as a function of the number of sampled Li ions. (e) Li-ion displacement distributions within 20 ps for the 13 and 52 Å cells. (f) Fine scan of the fitting-window dependence of *ln D* at 300 K.

Finite sampling additionally controls the statistical precision of the transport estimate. For a hopping process, the relative uncertainty decreases approximately with the number of statistically independent trajectories and hopping events,

$$\frac{\delta D}{D} \propto \left(N_{\text{eff}} N_{\text{hop}}\right)^{-1/2} \qquad (3)$$

Consistent with this scaling, the dispersion in $\ln D$ decreases substantially as the number of sampled Li ions increases (**Fig. 1d**), while the displacement distributions become smoother with improved sampling (**Fig. 1e**). Independent replicas therefore reduce statistical uncertainty, although they cannot remove the systematic effects associated with a restricted periodic cell.

A finer temporal scan at 300 K reveals a nonmonotonic variation of the apparent diffusivity over the first 10 ps, with the estimate remaining time dependent at the end of this interval (**Fig. 1f**). The approach to a plateau is observed on the longer timescale shown in **Fig. 1c**. This contrast motivates examining the dynamical regimes sampled by different analysis windows. Finite trajectory length, finite cell size, and limited statistics can therefore all alter the predicted ionic conductivity, but through physically distinct mechanisms. Their origins are resolved below before establishing a quantitative criterion for transport reliability.

*2.2 Distinct Dynamical Origins of Finite-Size and Finite-Time Errors*

The size and time dependences in **Fig. 1** originate from different dynamical mechanisms. In a finite periodic cell, an ion encounters periodically repeated environments over a shorter spatial scale, which can preserve dynamical correlations that would decay more completely in a larger system (**Fig. 2a**)[6, 13, 14, 17]. This behavior is reflected in the velocity autocorrelation function (VACF). The diffusion coefficient can equivalently be written through the Green-Kubo relation,

$$D = \frac{1}{d}\int_0^{\infty} C_{vv}(t)\,dt, \qquad (4)$$

where $C_{vv}(t) = \langle v(0)\cdot v(t)\rangle$[21]. In the present LLZO simulations, the VACF oscillations persist more strongly in the 13 Å cell than in the 52 Å cell, delaying cancellation between positive and negative contributions to the running integral (**Fig. 2b**). This provides a dynamical origin for the larger apparent diffusivity obtained from the smaller cell.

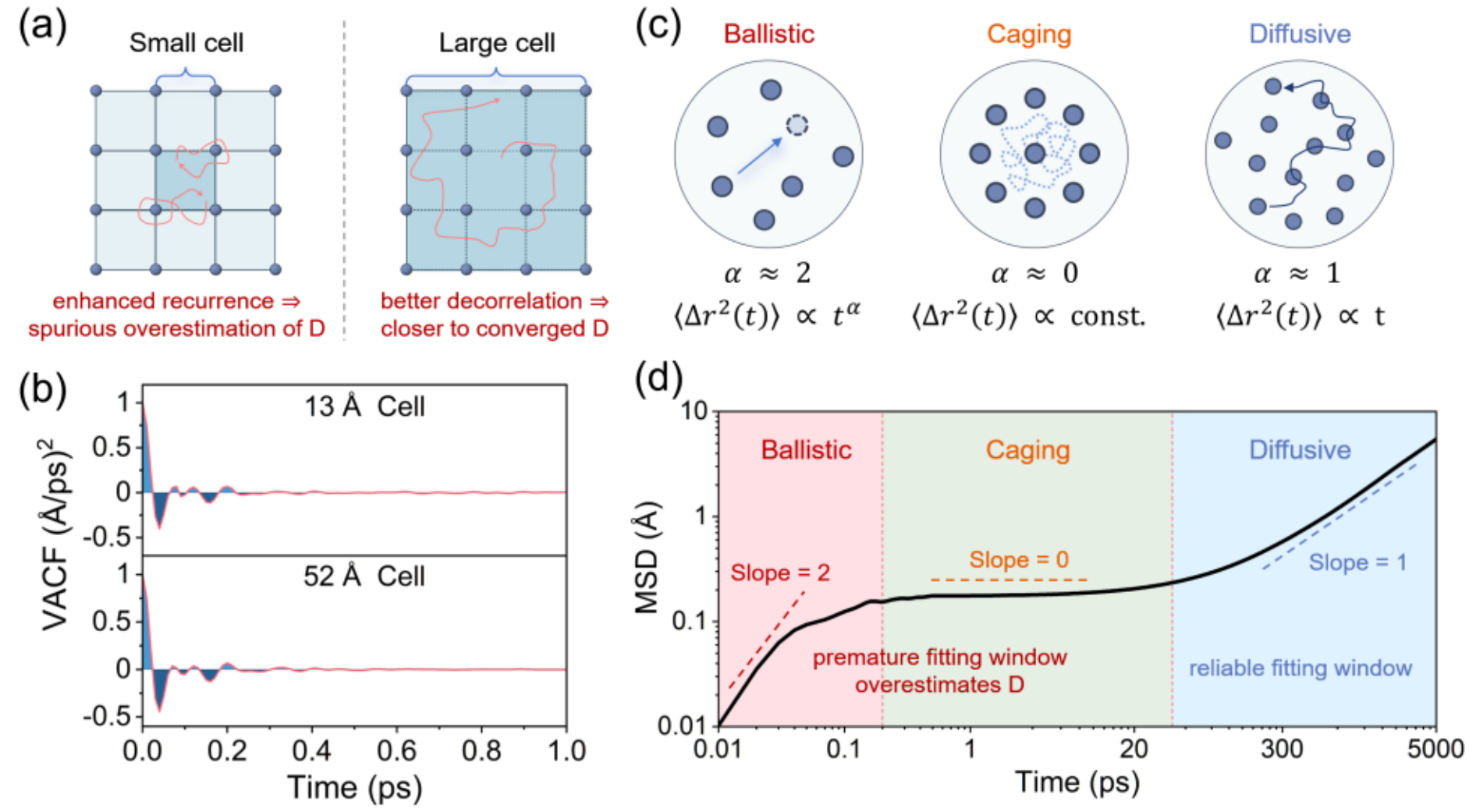

**Fig 2. Distinct dynamical origins of finite-size and finite-time biases.** (a) Schematic Li-ion trajectories in small and large periodic cells. (b) Li-ion VACFs for the 13 and 52 Å cell. (c,d) Three MSD regimes: ballistic, caging, and diffusive.

Periodic confinement also discretizes the accessible wavevectors,

$$k = \frac{2\pi}{L} n,\ k_{\min} = \frac{2\pi}{L}, \quad (5)$$

so that reducing $L$ removes progressively more long-wavelength modes[17]. The resulting redistribution of low-$k$ dynamics can prolong velocity correlations and enhance recurrence in finite systems. Importantly, this finite-size effect is distinct from statistical uncertainty: increasing the number of independent replicas improves sampling but does not change the spectrum imposed by the periodic cell.

Finite-time errors arise from a different origin. Before long-range diffusion is established, ionic motion evolves through ballistic, transiently confined, and diffusive regimes (**Fig. 2c,d**). These regimes are conveniently identified through the local MSD exponent,

$$\alpha(t) = \frac{d \ln [\mathrm{MSD}(t)]}{d \ln t}, \quad (6)$$

for which $\alpha \approx 2$, $\alpha \approx 0$, and $\alpha \approx 1$ characterize ballistic, caged, and normal diffusive motion, respectively[18-20]. A fitting window that overlaps the ballistic or crossover regime can therefore yield an excessive apparent slope, whereas a window dominated by caging does not represent long-range transport at all. The nonmonotonic behavior in **Fig. 1f** is consistent with contributions from different dynamical regimes, although its statistical component must be assessed independently.

Finite-size and finite-time errors thus have distinct microscopic origins, but both can produce apparently plausible transport coefficients before the relevant asymptotic limit is reached. The practical challenge is therefore to determine, directly from a finite trajectory, whether the sampled dynamics are sufficiently close to normal diffusion for the resulting ionic conductivity to be quantitatively reliable. The local exponent in **Eq. 6** provides the basis for this assessment, and its quantitative connection to transport error is established below.

*2.3 Local Dynamics Reveal Hidden Errors in Arrhenius Conductivity Prediction*

The consequences of incomplete dynamical convergence become particularly important when ionic conductivity is inferred through Arrhenius extrapolation. For thermally activated transport, the diffusivity is commonly described as

$$D(T) = D_0 \exp\left(-\frac{E_{\mathrm{a}}}{k_{\mathrm{B}} T}\right), \quad (7)$$

so that errors in individual diffusivities can propagate into both the fitted activation energy and the extrapolated low-temperature conductivity. **Fig. 3a** illustrates this challenge. At 1000-1600 K, frequent hopping produces substantial net Li displacement within several hundred picoseconds, whereas the 300 K trajectory requires a much longer time to reach comparable long-range motion. Similar simulation durations therefore do not correspond to equivalent levels of dynamical convergence at different temperatures.

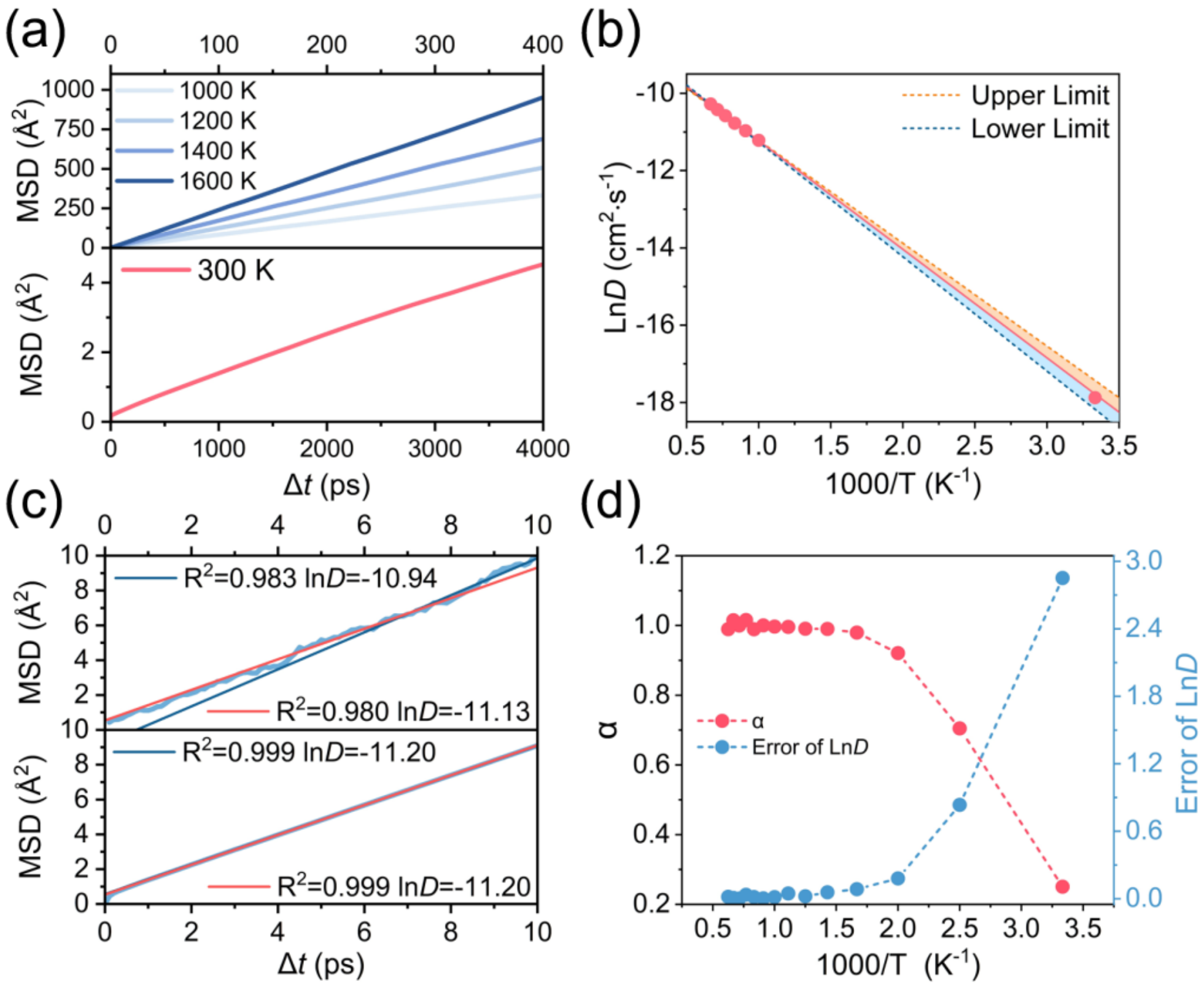

**Fig 3. Local dynamical convergence and the reliability of Arrhenius transport prediction.** (a) Li-ion MSD at 1000-1600 K and 300 K. (b) Arrhenius relation obtained from the calculated diffusivities; the upper and lower limits represent the variation associated with fitting-window selection. (c) Representative MSD fits showing that similarly high $R^2$ values can yield different $\ln D$. (d) Local MSD exponent $\alpha$and the corresponding error in ln D as functions of inverse temperature.

Despite this difference, the calculated diffusivities can still form an apparently well-behaved Arrhenius relation (**Fig. 3b**). The uncertainty associated with fitting-window selection, however, increases markedly toward lower temperature. More directly, different MSD intervals can exhibit similarly high coefficients of determination while yielding different diffusivities (**Fig. 3c**). A high $R^2$ therefore establishes only the local linearity of the selected MSD segment; it does not demonstrate that the slope has reached its asymptotic diffusive value. Consequently, an apparently excellent Arrhenius fit can conceal unresolved transport and propagate this error into the predicted ionic conductivity.

To quantify this effect, we define the logarithmic error of the apparent diffusivity as

$$\Delta \ln D = \ln D_{\mathrm{app}} - \ln D_{\mathrm{ref}}, \qquad (8)$$

where $D_{\mathrm{ref}}$ is obtained from the longest dynamically converged trajectory. **Fig. 3d** shows a clear correspondence between this error and the local MSD exponent. At high temperatures, $\alpha$ remains close to unity and $\Delta \ln D$ is small. As temperature decreases and the sampled trajectory becomes increasingly influenced by pre-diffusive dynamics, $\alpha$ departs from unity while the diffusivity error grows. The local exponent therefore contains dynamical information that conventional fitting statistics do not capture.

Importantly, $\alpha \approx 1$ should be regarded as a necessary dynamical condition rather than a complete convergence criterion, because statistical uncertainty and finite-size effects must still be assessed independently. Its value lies in determining whether a selected trajectory has reached a sufficiently diffusive regime for quantitative transport analysis. The next section establishes the explicit relation between the deviation of $\alpha$ from unity, the resulting error in ionic conductivity, and the minimum simulation time required to meet a prescribed accuracy.

*2.4 The Local MSD Exponent Quantifies Conductivity Error and Required Simulation Time*

The correspondence between the local MSD exponent and the diffusivity error enables dynamical convergence to be converted into a quantitative accuracy criterion. After the ballistic regime has decayed, but while a residual caging contribution remains, the MSD can be approximated as

$$\mathrm{MSD}(t) = U + 2dDt, \qquad (9)$$

where $U$ is the cage-plateau contribution and $D$ is the asymptotic diffusivity. The corresponding local exponent is

$$\alpha(t) = \frac{d\ln[\mathrm{MSD}(t)]}{d\ln t} = \frac{2dDt}{U + 2dDt}. \qquad (10)$$

If the diffusivity is estimated from the finite-time MSD, $D_{\mathrm{app}} = \mathrm{MSD}(t)/(2dt)$, **Eqs. 9** and **Eqs. 10** give

$$\frac{D_{\mathrm{app}}}{D} = \frac{1}{\alpha}, \; \frac{D_{\mathrm{app}} - D}{D} = \frac{1-\alpha}{\alpha}, \; \Delta \ln D = -\ln \alpha. \qquad (11)$$

The LLZO trajectories closely follow this relation (**Fig. 4a**), showing that the departure of $\alpha$ from unity provides a direct measure of the positive finite-time error in the caging-to-diffusion regime. For a Nernst-Einstein conductivity evaluated at fixed carrier concentration and temperature, the same relative error is transferred directly from diffusivity to conductivity,

$$\frac{\Delta\sigma_{\mathrm{NE}}}{\sigma_{\mathrm{NE}}} = \frac{\Delta D}{D} = \frac{1-\alpha}{\alpha}. \qquad (12)$$

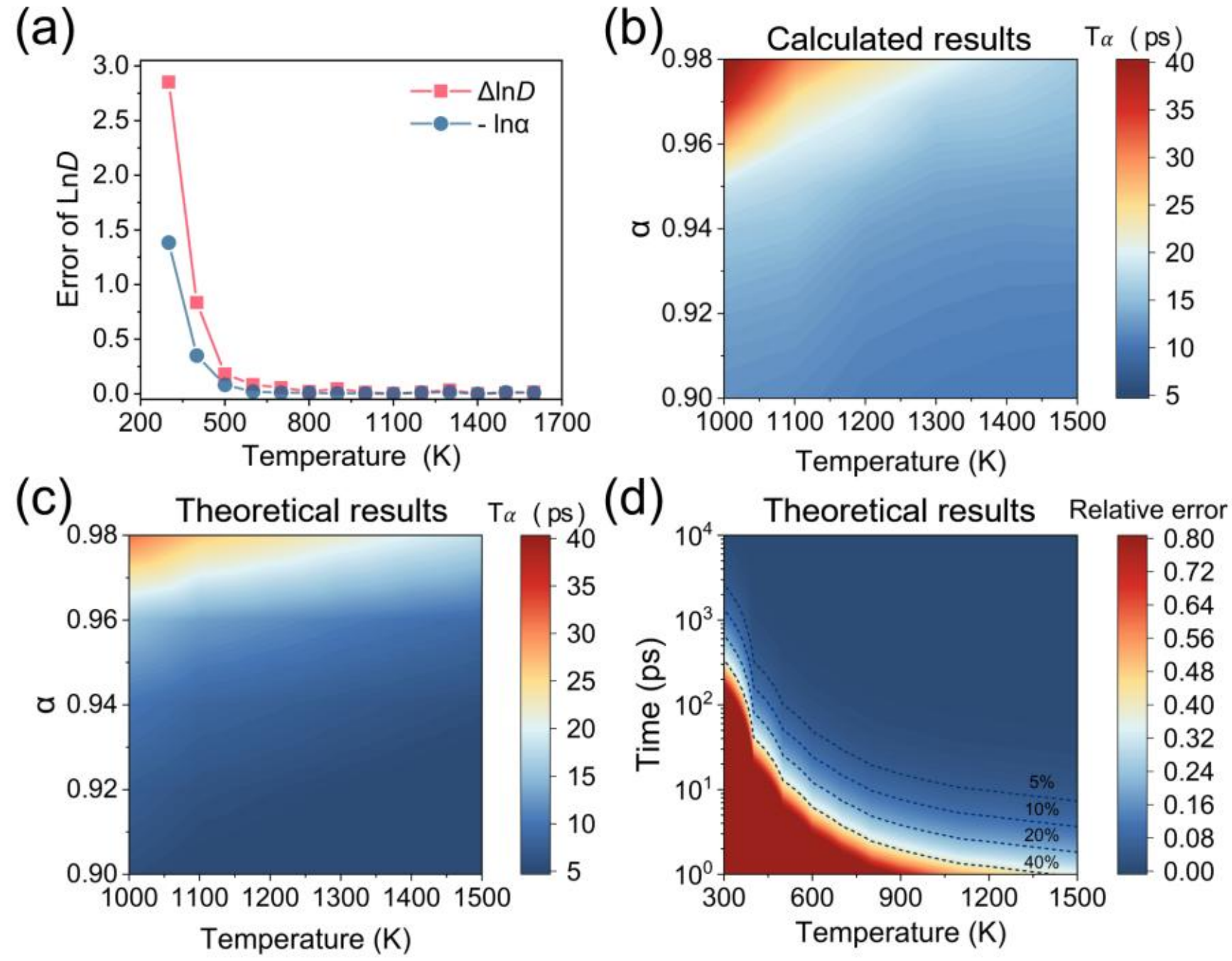

**Fig 4. Quantitative relation between dynamical convergence, transport error, and required simulation time.** (a) Temperature dependence of $\Delta \ln D$ and $-\ln \alpha$. (b) Minimum time $t_{\alpha}$ required for the LLZO trajectories to reach a prescribed local exponent. (c) Theoretical $t_{\alpha}$ predicted by the caging-diffusion model. (d) Predicted relative error as a function of temperature and simulation time; dashed contours denote errors of 5%, 10%, 20%, and 40%.

The local dynamical state can therefore be translated directly into an accuracy requirement for the predicted ionic conductivity. More importantly, this relation determines how long a simulation must be run to achieve a prescribed accuracy. For a target maximum relative error $\varepsilon$, **Eqs. 9-12** give

$$t_{\min} = \frac{U}{2dD\varepsilon} = \frac{U\alpha^*}{2dD(1-\alpha^*)}, \tag{13}$$

where $\alpha^* = 1/(1+\varepsilon)$ is the corresponding target exponent. The trajectory-derived minimum times increase rapidly as the target $\alpha$approaches unity and as temperature decreases (**Fig. 4b**). The same behavior is reproduced by the caging-diffusion model (**Fig. 4c**), indicating that the dominant temperature and accuracy dependences are captured by **Eq. 13**.

The strong temperature dependence follows naturally from the decrease in diffusivity. Combining Arrhenius transport, $D = D_0 \exp\left[-E_{\mathrm{a}}/(k_{\mathrm{B}}T)\right]$, with the harmonic scaling of the cage amplitude, $U \propto T/\omega^2$, gives

$$t_{\min} \propto \frac{T}{\omega^2 D_0 \varepsilon} \exp\left(\frac{E_{\mathrm{a}}}{k_{\mathrm{B}}T}\right) \tag{14}$$

The relative-error map in **Fig. 4d** summarizes this dependence: the simulation time required for a given accuracy rises steeply on cooling, and the 5%, 10%, 20%, and 40% contours define directly the computational effort associated with different error tolerances. Rather than assigning an arbitrary trajectory length to every temperature, **Eq. 13** allows the simulation time to be selected from the required accuracy of the ionic conductivity. This converts dynamical convergence from a qualitative check into a quantitative criterion for determining when a calculated transport property is sufficiently reliable for materials comparison and screening.

*2.5 Improving Conductivity Reliability under Restricted Computational Budgets*

The preceding analysis provides a criterion for identifying insufficient temporal sampling, but satisfying both the required trajectory length and full spatial convergence may be computationally demanding, particularly for ab initio molecular dynamics. We therefore examine two complementary strategies that address different sources of uncertainty: independent replicas for improving statistical sampling and selective single-axis expansion for probing finite-size effects.

For $n$ statistically independent replicas, the ensemble-averaged MSD is

$$\mathrm{M\bar{S}D}_n\,(t) = \frac{1}{n}\sum_{j=1}^{n} \mathrm{MSD}_j\,(t), \qquad (15)$$

and its statistical uncertainty decreases approximately as $n^{-1/2}$. Individual trajectories of the 13 Å cell exhibit appreciable fluctuations, whereas averaging eight independent simulations produces substantially smoother MSDs and more stable slopes (**Fig. 5a,b**). Consistently, pooling eight small-cell trajectories improves the displacement statistics and reproduces the main features of the distribution obtained from the 52 Å cell (**Fig. 5c**). Increasing the number of replicas from 1 to 4 and 8 also progressively narrows the uncertainty of the Arrhenius diffusivities (**Fig. 5d**). Replica averaging is therefore an efficient and readily parallelizable route to reducing statistical uncertainty. It does not, however, remove systematic finite-size effects because every replica retains the same periodically repeated spatial environment.

To probe spatial convergence at lower cost than isotropic cell expansion, we enlarge only one crystallographic direction and evaluate the corresponding directional diffusivity,

$$D_x = \frac{1}{2}\lim_{t\to\infty}\frac{d\langle \Delta x^2(t)\rangle}{dt}. \qquad (16)$$

Expansion of the 13 Å cell by a factor of four along $x$reduces the directional diffusivity and yields a more regular Arrhenius dependence (**Fig. 5e**), indicating that a substantial part of the small-cell enhancement originates from periodic confinement. The observed size response can be represented by

$$R_x(T) = \frac{D_x^{4\times1\times1}(T)}{D_x^{1\times1\times1}(T)}, \qquad (17)$$

which provides a practical estimate for correcting the scalar diffusivity of the original cubic cell,

$$D_{\text{corr}}(T) = R_x(T) D^{1\times1\times1}(T). \qquad (18)$$

The corrected values move toward the converged large-cell Arrhenius trend (**Fig. 5f**). This treatment relies on the approximate isotropy of cubic LLZO and should therefore be regarded as a computationally economical estimate of the dominant size dependence rather than a universal finite-size correction. For anisotropic solid electrolytes, convergence must be examined independently along the relevant transport directions.

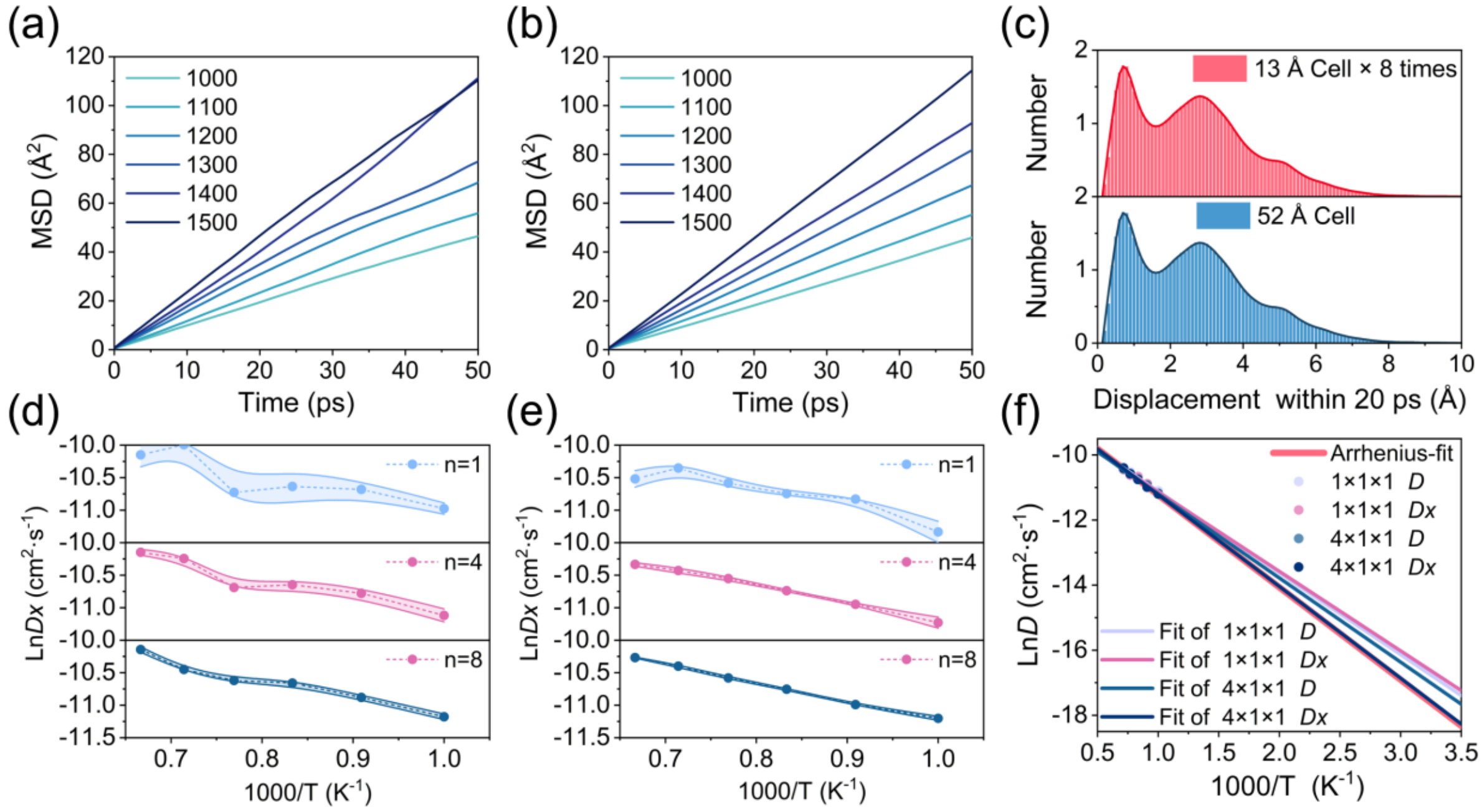

**Fig 5. Computationally efficient strategies for improving the reliability of ionic-transport predictions.** (a) MSD from the $L = 13$ Å cell at different temperatures. (b) MSD from eight independent simulations of the same $L = 13$ Å cell at different temperatures. (c) Number of ions that traverse beyond a fixed displacement within 20 ps. (d) Arrhenius dependence of $D_x$ obtained from 1, 4, and 8 replicas of the original 13 Å cell. (e) Corresponding $D_x$ after single-axis expansion. (f) Comparison of scalar and directional diffusivities for the $1 \times 1 \times 1$ and $4 \times 1 \times 1$ cells with the reference Arrhenius relation.

Independent replicas and spatial expansion thus address fundamentally different limitations: the former reduce statistical uncertainty, whereas the latter probe systematic size dependence. Combined with the $\alpha$-based time criterion, these results provide a practical basis for deciding whether additional computational effort should be devoted to longer trajectories, greater independent sampling, or larger simulation dimensions. Such targeted allocation of computational resources is particularly important when reliable ionic conductivities are required for large-scale comparison and screening of solid electrolytes.

## 3. CONCLUSION

Reliable prediction of ionic conductivity from molecular dynamics requires more than an apparently linear MSD or a well-fitted Arrhenius relation. Finite trajectory length, restricted cell size, and insufficient statistical sampling can each alter the calculated transport coefficient while leaving conventional indicators apparently well behaved. Their physical origins are distinct and therefore require different convergence checks.

Using LLZO as a representative solid electrolyte, we establish the local MSD exponent as a quantitative criterion for assessing finite-time transport reliability. In the caging-to-diffusion regime, the deviation of $\alpha$ from unity directly determines the error in the apparent diffusivity and, under the Nernst–Einstein approximation, the corresponding ionic conductivity. This relation further converts a prescribed transport accuracy into the minimum trajectory length required at a given temperature. Statistical uncertainty can be reduced efficiently through independent replicas, whereas finite-size effects require explicit spatial convergence or targeted cell expansion.

These results shift the assessment of simulated ionic conductivity from qualitative inspection to quantitative validation. Determining the reliability of transport coefficients before using them to rank candidate materials can prevent simulation-induced misclassification, focus computational and experimental resources on genuinely promising electrolytes, and ultimately enable more efficient development of solid electrolytes for solid-state batteries.

## 4. METHODS

Molecular dynamics simulations were performed for cubic LLZO using a previously validated machine-learning interatomic potential for LLZO[22-26]. Periodic cubic supercells with side lengths of approximately 13, 26, 39, 52, 65, and 78 Å were constructed to examine finite-size effects. All simulations were carried out in the canonical (NVT) ensemble under periodic boundary conditions with a time step of 1 fs. The models were equilibrated at each target temperature before production trajectories were collected. Simulations covered 300 K and the elevated-temperature range used for Arrhenius analysis. Additional calculations were performed by expanding the smallest cell fourfold along one crystallographic direction to evaluate directional finite-size effects.